# Information of income position and its impact on perceived tax burden and preference for redistribution: An Internet Survey Experiment

Eiji Yamamura

Seinan Gakuin University, Japan

Abstract

A customized internet survey experiment is conducted in Japan to examine how individuals' relative income position influences preferences for income redistribution and individual perceptions regarding income tax burden. I first asked respondents about their perceived income position in their country and their preferences for redistribution and perceived tax burden. In the follow-up survey for the treatment group, I provided information on their true income position and asked the same questions as in the first survey. For the control group, I did not provide their true income position and asked the same questions. I gathered a large sample that comprised observations of the treatment group (4,682) and the control group (2,268). The key findings suggest that after being informed of individuals' real income position, (1) individuals who thought their income position was higher than the true one perceived their tax burden to be larger, (2) individuals' preference for redistribution hardly changes, and (3) irreciprocal individuals perceive their tax burden to be larger and are more likely to prefer redistribution. However, the share of irreciprocal ones is small. This leads Japan to be a non-welfare state.

**Keywords:** Income position, Internet experiment; Perceived tax burden; Preference for redistribution

**JEL Classification:** H29; D63; D78; C99

# 1. Introduction

The traditional economic models indicate that relatively higher income persons benefit less from income redistribution, and thereby, do not prefer the redistribution (Romer, 1975; Meltzer and Richard, 1981). These models are based on an implicit assumption that persons know their income position in society[1]. Various works express doubt on the premise and examine its validity. Cruces et al. (2013) examine the impact of biased perception regarding individuals' income position on preference for redistribution through a survey experiment in Argentina[2]. They found that individuals who thought their income position higher than their true position become more likely to prefer redistribution if they are informed of their true position. Karadja et al. (2017) also investigate by liking the survey data to the administrative records of Sweden. Their key finding is that individuals who thought they are in a lower income position than their true one become less likely to prefer redistribution after being informed of their true relative income[3].

The quasi-experimental analysis found that the decision to protest tax depends on the private benefits and costs (Nathan et al., 2020). Gemmell et al. (2004) focused on tax preference and perceived tax cost to find that misperception about tax liabilities distorted individuals' tax preference. Therefore, private cost and benefit are perceived differently according to the situation and individual characteristics. Researchers consider individuals' perceptions and social norms when the government's policies about tax and public expenditure are analyzed (Gemmell et al., 2003; Feld and Larsen, 2012). Yamamura (2014) found that trust in government was correlated with a preference for redistribution and perceived tax burden. This study applied an experimental method to analyze how the provision of real income position influences perceived tax burden and preference for redistribution.

---

[1] Apart from the current income position, theoretical models indicate that expectations of upward and downward mobility determine individual attitudes toward redistribution (Piketty, 1995; Bénabou and Ok, 2001). This hypothesis is empirically supported by the works of Alesina and La Ferrara (2005) and Ravallion and Lokshin (2000).
[2] Respondents are informed about tax and distribution of income wealth (Kuziemko et al., 2015) and how other individuals perceive income distribution (Card et al., 2012).
[3] Providing true information changes people's views concerning various issues. For instance, people who have negative views on immigration before providing them information about immigrants can be more likely to support immigration if their misperceptions about the immigrants' characteristics are corrected (Grigorieff et al., 2020).

Accordingly, this study conducted an internet survey incorporating an experiment that is increasingly employed by researchers (e.g., Benjamin et al., 2014; Kuzemko et al., 2015; Fisman et al., 2020). The internet survey comprised a sample size of over 6,000, gathered by the first and follow-up surveys. The sample size is advantageous as it is nearly six times larger than that of existing experimental studies that examine the effect of the provision of true income position (Cruces et al., 2013; Karadja et al., 2017). These existing studies assess the difference in preference of respondents of the treatment group from those of the control group only after providing the information. However, they did not directly examine how preference changed by the provision of the information. The advantage of this study is to compare the same individual's preference and perception before and after being informed of one's true income position.

In Japan, I use the first survey to gather the respondent's perceived income position, household income, perceived tax burden, and preference for redistribution. Further, the question about subjective views about reciprocity and competition are included in the questionnaire because the effect of the provision of real income position differs according to subjective views about society and the market (Karadja et al., 2017)[4]. Subsequently, I assess if the respondents perceive their position in the income distribution appropriately.

The second follow-up survey was conducted two weeks after the first. It was an experiment to randomly assign two-third of the respondents to the treatment group where respondents were acquainted with personalized information concerning their true relative position in the income distribution. Moreover, we asked the same questions in the first survey about the perceived tax burden and preference for redistribution. Other respondents were randomly assigned to a control group where they were asked the same questions without the correct information of their income position.

The survey was conducted in Japan because increased immigration reduces support redistribution in Sweden (Dahlberg et al., 2012)[5], consistent with the negative relationship between racial fractionalization and social welfare spending (Alesina and Glaeser, 2004, p.141). As Japan is a racially homogenous society, it is expected to show significant social welfare spending. However, the social welfare spending of Japan is far smaller than Sweden and almost corresponding to that of the U.S (Alesina and Glaeser, 2004, p.141). Apropos to social welfare spending, Japan is considered a unique non-welfare country.

The formation of preference for redistribution differs according to institutional and social settings (Corneo and Grüner, 2002)[6]. The share of total public social expenditure over GDP is high in Nordic and Continental European countries. Nordic countries are characterized by high trust and low corruption whereas, Continental European countries are attributed with low trust and high corruption (Algan et al., 2016, 862). Algan et al. (2016) indicate that in Nordic countries, civic people support high taxes because they are surrounded by trustworthy individuals[7]. Conversely, in Continental European countries, uncivil people expect to benefit from the welfare state without bearing their costs and therefore support the welfare state.

---

[4] Shared societal values influence individuals' perceptions regarding redistribution and the welfare state (e.g., Gordon, 1989; Alesina and Angeletos, 2005; Wenzel, 2004, 2005a, 2005b; Klor and Shayo, 2007, 2010; Luttens and Valfort, 2011; Yamamura 2012; 2014).
[5] Notably, Nekby and Pettersson-Lidbom (2017) provided evidence of no relation between ethnic diversity and preference for redistribution.
[6] Many studies assessed the determinants of preference for redistribution (e.g., Ravallion and Lokshin, 2000; Fong, 2001; Corneo and Grüner, 2002; Alesina and La Ferrara, 2005; Luttmer and Singhal, 2011; Yamamura, 2012).

Compared with the countries as above, Japan's share of total public social expenditure is distinctly lower, whereas the levels of trust and corruption are in mid-point between Continental European and Nordic countries. In this regard, Japan is similar to the U.S. Alesina and Glaeser (2004) indicate that the U.S. redistributive policies remarkably differ from Western Europe. This might be because the U.S. citizens are more likely to believe that poverty is not society's fault; rather, the poverty-stricken are lazy than in the Continental European and Nordic countries (Alesina et al., 2004, p.188). Interestingly, the U.S and Japan share similar beliefs that poverty is not society's fault (Alesina et al., 2004, p.188) and that the poverty-stricken are lazy (Alesina et al., 2004, p. 2013).

However, a question remains unanswered. History, culture, and social background are different between the U.S and Japan. Is there a hidden mechanism to make Japan a non-welfare state? This study indicates that the provision of information did not change the preference for redistribution at all. Although respondents who learn that they are relatively penurious than they thought perceived their tax burden to be higher than before learning. Irreciprocal respondents perceived their tax burden higher and became more likely to prefer redistribution after learning their true income position. However, the share of irreciprocal people is considerably low in Japan.

---

[7] Trust is positively correlated with the size of the welfare state (Bjornskov and Svendsen, 2013).

The remainder of the paper is organized as follows. In Section 2, I describe the method of experimental strategy. In Section 3, I provide an overview of the data and the descriptive statistics. Further, I explain the estimation approach in Section 4 and provide the estimation results and their interpretation in Section 5. The final section offers some conclusions and draws out some implications for future research.

## 2. Experimental Design Of Internet Survey And Data

2.1. The first survey

To gather a large sample and conduct the experiment, I independently collected individual-level data through internet surveys covering all parts of Japan. In the first survey, from October 25–30, 2018, I invited 9,130 subjects to participate. Subsequently, I gathered 7,855 observations, and the response rate was 88%. Regarding basic information, the respondents were asked about their age, residential prefecture, educational background, job status, marital status, and the number of children. As described in Table 1, various subjective views were asked, such as perceived tax burden and preference for redistribution, which were generally used in existing works (e.g., Corneo and Grüner, 2002; Ohtake and Tomioka, 2004; Alesina and La Ferrara, 2005; Alesina and Giuliano, 2009; Yamamura, 2012, 2014). In addition, the questionnaire includes a question about views regarding irreciprocity and competition. Its definition follows previous studies (Yamamura and Tsutsui, 2019; Ito et al., 2021). In the Appendix, the upper part of Figure A1 demonstrates a screen where the respondents see when they are asked about views.

After the abovementioned questions, they were asked to report their annual household income from the previous year and state their perceived position in society by answering the following question:

*"How many percent is your household income position located in Japanese society?"*

In the Appendix, the lower part of Figure A1 demonstrates a screen where the respondents see when they are asked about their household income position. There are ten choices with ten percentage ranges starting from "0–9% (the top)" and ending with "90–100% (the bottom)." In the estimation, the values are converted to "0–9% (the bottom)" and "90–100% (the top)," and its mid-point values are used as perceived income position in the estimation. Hence, the larger the mid-point value, the higher respondent's income position. Similarly, the respondents are expected to answer the question regarding their household income level by picking one of the 12 choices[8]. The mid-point value is used as household income in the analysis.

2.2. The follow-up survey

The second follow-up survey was conducted two weeks after the first survey. It was a randomized experiment where two-thirds of the respondents were informed of their true

---

[8] A total of 12 choices are suggested by 10,000 as a unit: (1) 0–99, (2) 100–199, (3) 200–399, (4) 400–599, (5) 600–799, (6) 800–999, (7)1000–1199, (8) 1200–1399, (9) 1400–1599, (10) 1600–1799, (11) 1800–1999, and (12) over 2000. The mid-point of the respondent's selected income group is used as the household income. For instance, household income is 150 for those who selected "100–199." In the exceptional is the top group, household income is defined as 2300 for those who selected "over 2000" because the mid-point is unknown. Those who selected "over 2000" is only 1.5 %; therefore, its effect is considered less significant.

income position in the question about preference for redistribution and perceived tax burden.

In the follow-up survey, I randomly assigned subjects into three groups: (A) any information about income position is not indicated in the questions, which occupied one-fifth of the subjects; (B) correct information of respondents' income position in the question, which occupied two-fifth of the subjects; (C) world rank of Japan in income inequality is indicated—which is planned to be used for different analysis—which occupied two-fifth of the subjects. To assess the effects of providing information on an individual's income position, Group (A) is the control group, and Group (B) is the treatment group in this study.

A total of 7,855 subjects participated in the first survey. I gathered 6,715 observations, and the response rate was 85% in the follow-up survey. These are divided into Group (A) (2,744), Group (B) (1,342), and Group (C) (2,628). Further, some respondents did not answer the questions used in the estimation, such as the number of children in the first survey. Eventually, individuals included in the estimation were reduced to 2,341 and 1,134 for Group (A) and Group (B), respectively. Details of the sampling method are provided in the Appendix.

The subjective well-being of people depends not only on one's income but also on the income level of neighboring people (e.g., Clark and Oswald, 1996; Luttmer, 2005; Card et al., 2012). Individuals' preference for redistribution becomes smaller as the average incomes of individuals' ethnic and religious groups are higher (Quattrociocchi, 2018). In considering income distribution, "individuals observe the income levels of no more than a sub-sample of the population and must then infer the entire distribution from that information" (Cruces et al., 2013, p. 101). An individual's income position depends on groups where individuals belong, and its effect possibly changed according to how the group is defined. For instance, individuals belong to a residential locality while they also belong to a cohort group. The effect of income position in one's locality is possibly different from that in one's cohort. Biases in the perception of one's own income position in society are "significantly correlated with the respondents' relative position within the reference group (as proxied by area of residence)" (Cruces et al., 2013, p. 101). Rather than real income position in the whole society, this study examines how respondents are influenced by correct information of the income position within different groups.

Using the data obtained from the first survey, individual income position is calculated in three ways before the follow-up survey: (1) income position within a residential locality (prefecture)[9]; (2) the position in the same cohort[10]; and (3) the position within the same final educational background. The mean value of income positions is 48.2, 47.3, and 48.3 in the residential locality, cohort, and educational group, respectively. Hence, on average, there is a small variation.

In the Appendix, the upper part of Figure A2 demonstrates a screen indicating the respondent's income positions within a residential locality (prefecture), the individual's position in the same cohort, and the position within the same final educational background. The three different measures of the position are colored in red, which were presented directly above the questions for the subjective views about perceived tax burden and preference for redistribution. Apart from the provision of an individual's income position, these questions were the same as in the first survey. Specifically, the respondents of the treatment group naturally answered the question after learning their income positions measured variously.

---

[9] An administrative subdivision called a prefecture in Japan. Localities are defined by 47 prefectures, which are generally used in previous works (e.g., Yamamura, 2012, 2014; Yamamura and Tsutsui, 2019).
[10] Cohort groups are defined by every 10 birth years.

However, even for the treatment group, the questionnaire did not indicate the respondent's own choice about one's subjective income position in society, perceived tax burden, and preference for redistribution in the first wave. As explained in subsection 2.3, in the estimation of biases of income position, I used the differences between three different objective income positions as mentioned above, and the subjective income position was obtained in the first survey. However, the respondents did not have the information about the difference in the follow-up survey, although they could consider it by recalling their own choice in the first survey. In the first survey, respondents were vaguely asked about their income position in society. Hence, the group in which they considered their position varied according to individuals. Therefore, using three different biases, we could check which groups they considered primarily for income position.

2.3. Data

As explained in subsection 2.2., 3,475 respondents were included in the estimation, of which 2,341 were a part of the treatment group and 1,134 a part of the control group. They respond to both the first and follow-up surveys, and therefore, the sub-sample size used in the estimations is 4,682 for the treatment group and 2,268 for the control group.

Table 1 reports the results of the mean difference test in the first wave. In most cases, there is no significant difference of various variables between the treatment and control groups. The exceptional cases are BIAS_AGE and POS_BIAS_AG, but its statistical difference is only at the 10 % level. Overall, respondents of the treatment group have almost the same characteristics and thus can be comparable. Apart from the dependent variables, such as TAX_BURDEN and PREF_REDIST, variables used in the estimation have the same value in both the first and follow-up surveys.

According to Dohmen et al. (2009), reciprocal people tend to work harder, have more friends, and show higher subjective well-being. Further, they are more willing to invest in a positive self-image by engaging in positive reciprocity. Therefore, reciprocal people are considered to have a sense of public morality. For the convenience of interpreting the results, I define irreciprocity as the degree of "not reciprocal" [11]. The mean of IRRECIPRO is considerably small—approximately 1.95 on the five-point scale—indicating that most respondents chose 1 or 2, conducive to reciprocal characteristics. Subsequently, I add its dummy (IRRECIPRO_D), where respondents chose 3, 4, or 5, which are not considered reciprocal. Its values are around 0.23; therefore, 23% of people are irreciprocal. As illustrated in Appendix Figure A3, most individuals are not irreciprocal.

A positive view about market competition is negatively related to preference for redistribution (Glaeser et al., 2004). Hence, I further explore the effects of their view about competition. The degree of agreeing with that benefit from the competition is captured by COMEPT. Appendix Figure A4 demonstrates that most people support the competition. As an alternative variable, I use its dummy (COMEPT_D), wherein the respondents chose 3, 4, or 5, indicating a positive view about competition.

---

[11] Dohmen et al. (2009) defined positive reciprocity by asking the degree of reciprocal behavior such as "If someone does me a favour, I am prepared to return it," while they defined negative reciprocity by asking different questions, such as "If I suffer a serious wrong, I will take revenge as soon as possible, no matter what the cost." There is only one question about "reciprocity," as shown in Table 1. The degree of irreciprocity is measured simply as the degree of disagreeing with the reciprocal behavior.

To verify the validity of the experiment, Figures 1–5 compare the distribution of key variables in the first survey between the treatment and control groups. Figure 1 compares the perceived tax burden between the treatment and control groups in the first survey. The respondents are unlikely to perceive the tax burden to be low. Its distribution of the treatment group is highly similar to that of the control group. Figure 2 compares the preference for redistribution and demonstrates results similar to those in Figure 3. There are no differences in dependent variables before informing the correct information of the income position for the treatment group. Arguably, people who perceived the tax burden to be larger are more likely to prefer redistribution even before they know their real income position.

Using Kernel density, Figure 3 shows that the distribution of household income skewed toward the left, and no difference between the groups. In Figure 4, there is a similar distribution of perceived income position between the treatment and control groups, although it is slightly skewed toward the right. Figures 5 (a)–(c) illustrated the bias of income positions in three ways. The shape of distributions is unlikely to skew, and its peak is around 0. Respondents underestimated that their income position has a negative value, which is considered as negative bias. Conversely, the respondents who overestimated their income position have a positive value, which is considered as positive bias. The share of positive bias is on par with a negative bias, which documents that misperceptions about relative income are balanced. This finding is similarly observed in Argentina (Cruces et al., 2013), Spain (Fernández-Albertos and Kuo, 2015), and Germany (Engelhardt and Wagener, 2016) but varies in Sweden, where most people show negative bias (Karadja et al., 2017).

Considering Table 1, Figures 1–5 verified the validity of the experimental analysis using the online survey.

## 3. The Econometric Model

Following Karadja et al. (2017), I assess how the bias of income position occurs using the simple OLS model as the following specification. Ensuingly, I conducted the Fixed Effects model in the framework of the difference-in-difference method. The purpose was to examine the effect of providing the correct information of income position on the perceived tax burden and the preference for redistribution by including a cross-term between the follow-up survey dummy and the biases of income position. The specification of estimation model is as follows:

TAX_BURDEN (or PREF_REDIST )$_{it}$ = $\alpha_1$ SECOND $_t$ *BIAS_LOCAL $_i$ (BIAS_AGE $_i$, or BIAS_EDU $_i$ )+ $\alpha_2$ SECOND $_t$ *INCOM $_i$ + $\alpha_2$ SECOND $_t$ *GINI $_i$ + $\alpha_2$ SECOND $_t$ *AGE $_i$ +$\alpha_2$ SECOND $_t$ *CHILD $_i$ + $\alpha_2$ SECOND $_t$+  m $_i$ + u$_{it}$,

The dependent variable TAX_BURDEN (or PREF_REDIST )$_{it}$ is individual *i*'s perceived tax burden (or preference for redistribution) at the time of survey *t*. SECOND is the follow-up survey dummy. BIAS_LOCAL is the bias of income position for individual *i*, which does not change in the first and second surveys. The key independent variable is SECOND *BIAS_LOCAL, which is the interaction term between SECOND and BIAS_LOCAL. Instead, for robustness check, BIAS_AGE $_i$ (or BIAS_EDU $_i$ ) is used to interact with SECOND in alternative specifications. I could do a robustness check and additionally compare the importance between the three groups when the respondents considered their income position. m $_i$ is the individual's time invariant fixed effects. $u_{it}$ is an error term. For control variables, there are various interaction terms between SECOND and economic (or demographic) variables.

I conducted estimations using a sub-sample of the treatment and control groups separately. There were only two weeks between the first and follow-up surveys; therefore, most of the individual's factors did not change, apart from the provision of the information only for the treatment group but not for the control group. SECOND *BIAS_LOCAL is included to examine the effect of providing the correct information of income position. In the treatment group, SECOND *BIAS_LOCAL has a positive sign if tax is perceived to be higher than their thought after learning that their real income position is lower. Further, in the model where PREF_REDIST is a dependent variable, SECOND *BIAS_LOCAL has the positive sign if the results are consistent with the previous works analyzing Argentina (Cruces et al., 2013) and Sweden (Karadja et al., 2017). Meanwhile, in the control group, the respondents did not know their real income position even in the follow-up survey. Particularly, the situation did not differ from the first survey, and they were asked to answer the same questions about perceived tax burden and preference for redistribution. Therefore, the coefficient of SECOND *BIAS_LOCAL is predicted to be statistically insignificant.

In the baseline model as above, the effects of positive and negative bias are not decomposed. For closely examining the effects of positive and negative bias separately, as conducted in Karadja et al. (2017), the interaction term SECOND *BIAS_LOCAL is decomposed into SECOND *POS_BIAS_LOCAL and SECOND *NEG_BIAS_LOCAL Further, the preference for redistribution is expected to depend on a sense of public morality (Algan et al., 2016). Moreover, the attitude toward competition also influences redistribution because an individual considering competition is unlikely to depend on the government. Karadja et al. (2017) examined how the effect of the information changed by the political partly preferences before learning real income position. However, in the experimental study, it has not been examined whether the effect of providing the information of real income position varies according to a sense of public morality and attitude toward the competition. In this paper, a sense of public morality is captured by reciprocity, such as the degree that individual returned to the benefactor if someone did the respondent a favor. For the convenience of interpretation, we use a proxy for opposite characteristics—irreciprocity. Hence, we added interaction terms between SECOND and IRRCIPRO (or COMPET) to the model. From this model, I can examine how and to what degree Japanese people's attitudes toward the welfare state differ from those in the Nordic and Continental European countries (Algan et al., 2016).

## 4. Estimation Results

### 4.1. Determinants of bias

Tables 2 show the results for examining the bias of an individual's income position based on the first survey. In columns (1)–(3), bias (dependent variable) has both positive and negative values because there are positive and negative biases. In addition, following the way of Karadja et al. (2017), the absolute value of bias is the dependent variable in columns (4)–(6) to examine the degree of inaccuracy of the respondent's evaluation of their income in society. IRRECIPRO shows the positive sign and statistical significance in all columns. Irreciprocal individuals tend to over-evaluate their income position, and their bias is larger. Conversely, COMPET indicates the negative sign and statistical significance in all columns.

HINCOM produces the significant negative sign in columns (1)–(3), which is consistent with the finding of Argentina that wealthier individuals underestimate their income rank (Cruces et al., 2013). Further, the significant negative sign is also observed in columns (4)–(6), indicating that wealthier people's bias in the income position is smaller. Respondents who had a positive view about competition are likely to under-evaluate their income position, but their evaluation is more likely to be accurate. UNIV, a higher degree of educational background, shows the significant negative sign in columns (4)–(6), implying that educated people tend to accurately evaluate their income position, which is consistent with Karadja et al. (2017). In my interpretation, the cost for acquiring information about their income position is lower for wealthier and highly educated individuals. In this regard, the mechanism of having bias in Japan is similar to other countries (Cruces et al., 2013; Karadja et al., 2017).

TREAT_DUMMY does not indicate statistical significance in any columns, implying no difference of biases between the treatment and control groups. Hence, consistent with Table 1 and Figures 1–5, the validity of the experiment is verified. Specifically, respondents are randomly assigned to treatment and control groups.

Appendix Table A2 reports the results when the dependent variable perceives their income position. The reported results are limited to key variables such as objective income position in the whole sample and income position in various groups, although various control groups are included in the table. The results are comparable to those of Cruces et al. (2013). Only subjective income position in the same educational background shows statistical significance. Further, its sign of the coefficient is negative, determining that respondents tend to underestimate their income. This can be interpreted as follows: the percentage of who think the poverty-stricken are lazy is approximately 60% in the U.S. and Japan, which is remarkably higher than in Nordic and Continental European countries (Alesina and Glaeser, 2004, p.211). Especially within the same educational background group, lower income reflects laziness, which results in poverty-stricken individuals feeling shameful in admitting their real income position. Meanwhile, those who earn a higher income possibly become an objective of jealousy by others with the same educational background. Wealthier people with the same educational level denigrate others, resulting in conflict between them. Naturally, wealthier respondents may avoid reporting their income position higher than those with the same educational background. In Japan, people tend to constitute the relevant reference group for the formation of the perception of income distribution. Concerning the reference group, Japanese are more likely to consider the group with the same educational background much more than those of residential locality and the same cohort. This is inconsistent with the finding of Cruces et al. (2013), demonstrating the positive correlation between objective income position within a residential locality and perceived income position.

4.2. Estimation results with perceived tax burden as a dependent variable.

Tables 3–6 report the results of estimation about perceived tax burden using the Fixed effects model. First, Table 3 shows the results of the baseline model. As for results based on the sub-sample of the treatment group in columns (1), (3), and (5), we observed the positive sign and its statistical significance at the 1% level in key variables SECOND*BIAS_LOCAL, SECOND*BIAS_AGE, and SECOND*BIAS_EDU. Comparison with members of the same age group or educational background is more important than those with the residential locality. Meanwhile, these variables are not statistically significant in the results using the sub-sample of the control group in columns

(2), (4), and (6). Concerning the results of key variables, there is a robust and clear difference between the treatment and control groups. These results indicate that the provision of correct information changed the perceived tax burden. Moreover, statistical significance is hardly observed in the results of other control variables.

In Tables 4, 5, and 6, only key variables are reported, although the set of control variables shown in Table 2 are included in the model. The results of Table 4 for the treatment group determine that the interaction terms between SECOND and the variable of positive bias show a positive sign and is statistically significant at the 1% level. However, the terms between SECOND and the variable of negative bias do not show statistical significance in any columns. Absolute values of coefficients are around 0.39, 0.31, and 0.36 for SECOND*POS_BIAS_LOCAL, SECOND*POS_ BIAS_AGE, and SECOND*POS_BIAS_EDU, respectively. A 1% increase of the positive bias enables people to perceive tax by 0.31 to 0.39 points on the 5 point scale. This is a fairly sizable effect. Thus, people emphasize the importance of the group of the same residential locality than the same age group and educational background when they compare their income to others. Consistent with Table 3, the statistical significance of interaction terms in any results for the control group is not observed. The combined results of Tables 3 and 4 reveal that individuals who assumed that their income position was higher than the true one perceived their tax burden to be more significant after knowing their real income position.

Table 5 reports the results of interaction terms between SECOND and the individual's view prior to the treatment. The significant positive sign of SECOND*IRRECIPRO and SECOND*IRRECIPRO_D is observed for results using a sub-sample of the treatment group but not for the control group. The absolute value of the coefficient is 0.211 for SECOND*IRRECIPRO_D, implying that the irrecipocal people perceive tax burden to be larger by 0.21 points on the five-point scales after being informed of the real income position than before it. Various interaction terms are controlled. Therefore, the results do not reflect the change of effects of objective economic conditions such as income level and GINI in the residential locality after providing the information. Hence, in my interpretation, the provision of true income position leads to uncivil irreciprocal individuals perceiving their burden of tax to be more significant. "If taxpayers are reciprocal, they are more willing to pay taxes if the tax system is considered fair and if other taxpayers are expected to pay their taxes as well" (Dohmen et al., 2009, p.610). Derived from it, irreciprocal people perceive tax to be larger if their income position is different from what they thought because they assume that others did not report the real income to evade tax.

The information effects possibly vary according to whether the bias is positive or negative. To explore it, Table 6 indicates results using a sub-sample of negative and positive biases separately. Panel A of Table 6 shows the significant positive sign of SECOND*IRRECIPRO and SECOND*IRRECIPRO_D for results using the sub-sample of positive bias but also for negative bias. One interpretation is that the perceived bias of income position leads irreciprocal people to believe that the tax system is considered unfair. Moreover, the negative sign of SECOND*COMPET is statistically significant, whereas that of SECOND*COMPET_D is not significant. Hence, the competition did not influence the perceived tax burden. Supporting the validity of the experiment, Panel B showing results for the control group did not indicate statistical significance in any results.

4.3. Estimation results where preference for redistribution is a dependent variable

Further, we examine the effects of the correct information on preference for redistribution. Table 7 reports the results of the baseline model. Inconsistent with the results of the perceived tax burden in Table 3, key interaction terms between SECOND and the bias of income position are not statistically significant in any columns. Surprisingly, learning the individuals' true income position did not influence their preference for redistribution in any case. This is remarkably different from the previous experimental works in Argentina (Cruces et al., 2013) and Sweden (Karadja et al., 2017).

As shown in Table 8, even after decomposing bias into positive and negative ones, any interaction terms between SECOND and biases were not statistically significant in any columns. The striking results are not in line with the traditional theoretical inference that a relatively higher income person benefits less from income redistribution and does not prefer the redistribution (Romer, 1975; Meltzer and Richard, 1981). However, respondents are not considered irrational because the results of Tables 3–4 imply that they rationally perceive their tax burden to be higher when they learn that their income position is lower than they thought.

I interpret these results jointly. As shown in Appendix Table A1, based on World Value Surveys, the rate of those who are confident in the government is about 24% in Japan, which is distinctly lower than in other countries such as Sweden (59%), Argentina (32%), and the U.S (32 %). Hence, the Japanese are less likely to be confident in the government. Existing literature provided evidence that trust and confidence in the government increase demand redistribution (Yamamura, 2014; Kuziemko et al., 2015). Therefore, the Japanese are not incentivized to depend on the government to demand income redistribution even if they know that it benefits them.

Concerning the results of preference for redistribution, there are some other possible interpretations. Similar to the U.S., the Japanese are likely to believe that the poverty-stricken are lazy (Alesina et al., 2004, p. 2013). Naturally, a beneficiary from income redistribution is likely to be labeled as an encumbrance for society or a social failure. The theoretical model of Bénabou and Tirole (2006) demonstrates that individuals gain utility from conforming to social expectations or morals. Hence, those who can benefit from redistribution have motivations to prefer redistribution to protect their human dignity (Banerjee and Duflo, 2019, Ch9). The individual is reluctant to disclose one's real intention even when one prefers redistribution after learning the real position. Another interpretation can be derived from the theoretical hypothesis (Piketty, 1995; Benabou and Ok, 2001). The Japanese expect that income will rise in the future even if they are currently poverty-stricken. However, in the real setting, the average growth rate of the real GDP between 1995–2018 was only 0.9% in Japan[12]. Therefore, generally, the Japanese are unlikely to expect a rise in their income in the future.

Turning to Table 9, a significant positive sign was observed for interaction terms of SECOND and proxy variable of the irreciprocity for the treatment group but not for the control group. The absolute value of the coefficient is 0.20 for SECOND*IRRECIPRO_D, implying that the irrecipocal individuals prefer redistribution by 0.20 points on the five-point scale after being informed of the real income position than before it. Therefore, uncivil people more likely prefer redistribution after being informed of their real income position. The SECOND * COMPET produced a significant negative sign for the treatment group, although a similar result is observed for the control

---

[12] Official website of Cabinet of office of Japan: https://www.esri.cao.go.jp/jp/sna/menu.html (accessed on June 9, 2021).

group. Conversely, SECOND * COMPET_D was not statistically significant at all.

In Panel A of Table 10, on the one hand, I observed a significant positive sign for both SECOND*IRREIPRO and SECOND*IRRECIPRO_D when the bias is positive. On the other hand, statistical significance is observed for SECOND*IRREIPRO but not for SECOND*IRRECIPRO_D when their bias is negative. In column (2) of Panel A, the absolute value of the coefficient is 0.24 for SECOND*IRRECIPRO_D, implying that irrecipocal individuals prefer redistribution by 0.24 points on the five-point scale after knowing that their real income position is lower than they thought. However, irreciprocal people's preference for redistribution did not change when they knew their real income position is higher than they thought. In Panel B of control group results, neither SECOND*IRREIPRO nor SECOND*IRRECIPRO_D are statistically significant. A possible interpretation is that after knowing the perceived bias of one's income position, irreciprocal individuals did not admit their misperception and believed that an unfair tax system caused bias. They naturally become more likely to prefer redistribution. These results showed that uncivil individuals who learn that their income position is lower than they thought tend to prefer redistribution. This is in line with the observation in the Continental European countries found by Algan et al. (2016). An alternative interpretation suggests that reciprocal individuals tend to work hard and do not prefer redistribution (Dohmen et al., 2009). My finding is consistent with the classical work of Corneo and Grüner (2002) and provided evidence that people who believe in the importance of personal hard work oppose redistribution.

Concerning SECOND*COMPET for the treatment group, it shows the negative sign and statistical significance when respondents have positive and negative biases. For the control group, the significant negative sign is observed only when respondents have a negative bias. Further, SECOND*COMPET _D was not statistically significant in any results. Hence, preference redistribution did not change according to the view regarding competition after learning the information of income position.

The observations thus far lead me to argue that the Japanese rationally perceived tax burden to be larger than before being informed of their real income position if their real income position is lower than they thought. However, the provision of the information did not change their preference for redistribution, which is remarkably different from the results reported in previous experimental studies in other countries. One of the interpretations is that the share of people confident in the government is considerably low—nearly 24%—and the Japanese do not prefer the government's income redistribution. Moreover, uncivil individuals become more likely to perceive tax burden to be larger and prefer redistribution after learning their real income position.

Algan et. al. (2018) argues that "the majority of the population is made up of untrustworthy individuals who exploit the advantages provided by the welfare state at the expense of a minority of trustworthy individuals" (Algan et. al., 2018, 882)[13]. In the case of Japan, the share of uncivil individuals is considerably low, around 24% (Table 1), and the demand for a welfare state is smaller than in Continental European countries. Otherwise, most people believe that the poverty-stricken are lazy or make a failure of their life. In society, "many people, therefore, resist having to admit to themselves or others that they are poverty-stricken enough to deserve help" (Banerjee and Duflo, 2019,

---

[13] The formation of the labor insurance institution depends on civic virtue (Algan and Cahuc, 2009).

p281). Accordingly, the Japanese are more likely to emphasize interpersonal reciprocity and market competition, which form a non-welfare state.

## 5. Conclusion

Japan is characterized by a non-welfare state and located in the mid-point between high and low trust and also between high and low corrupted government. Hence, the characteristics of Japan are different from those of welfare states such as Nordic and Continental European countries analyzed by Algan et al. (2016). At this point, Japan is more similar to the U.S. than the European countries. Alesina and Glaeser (2004) contrast the redistributive policies of the U.S. and Western Europe partly because Europeans appear to prefer more equal societies than the U.S. The question of how and why are individuals of non-welfare states are less inclined to demand to redistribute arises.

This study enabled me to scrutinize how the right information of income position change enables the preference for redistribution and perceived tax burden and how the effect of the correct information varies according to subjective characteristics. The finding of this study differs from the existing works suggesting that individuals pursue rational self-interest from policies of income redistribution if they learn their true income position (Cruces et al., 2013; Karadja et al., 2017). My finding is that an individual's preference for redistribution did not change despite knowing one's income position, whereas one's perceived tax burden becomes higher if one knows that their true income position is lower than they thought. I interpret it in the following ways. In Japan, the confidence in the government is remarkably lower than in other countries. Thus, people are less likely to demand redistribution. Another interpretation is that people are reluctant to disclose one's real intention even if they prefer redistribution because a beneficiary is labeled as an encumbrance for society or a social failure. They have motivations to avoid being considered lazy for protecting their human dignity (Banerjee and Duflo, 2019, Ch9). Meanwhile, individuals do not intend to conceal their perceived tax burden.

In addition, I found that irreciprocal respondents perceived their tax burden as higher and became more likely to prefer redistribution after learning their true income position. This is consistent with the argument that a decline of a sense of public morality leads the Japanese to demand an expansion of social security without increasing taxes (Kume et al., 2018). However, the share of uncivil individuals is considerably low, and most individuals have a sense of public morality. The combined findings of this study indicate why Japan is a non-welfare state.

I provide several speculations about the reasons that the provision of real income position does not change an individual's preference for redistribution. It is necessary to scrutinize the mechanism by a further experimental method and check the external validity by conducting the estimation in other non-welfare states. Moreover, the experimental method is difficult to be applied for long-term historical change of economic situation. Thus, a historical study is useful as a compliment for experimental study to see the overall picture. The share of the government's social welfare spending is at the same level between Continental Europe and the U.S. in 1870. However, the gap between them increased consistently until 1998 (Alesina and Glaeser, 2004, p.20). From a comparative viewpoint, it is necessary to assess the change of Japanese social welfare spending historically. In the future, social security expenditure is predicted to expand because of the coming unprecedented aging society (Kume et al., 2018). There is a possibility of a rise in populism in Japan. Further, not only the Japanese but also the citizens in various countries seem to confront the possibility that individuals support populism, which hampers social and economic sustainability. It is crucial to explore how a sense of civic morality is fostered in an aging society. These are the remaining issue to be addressed in future studies.

Acknowledgments

Funding

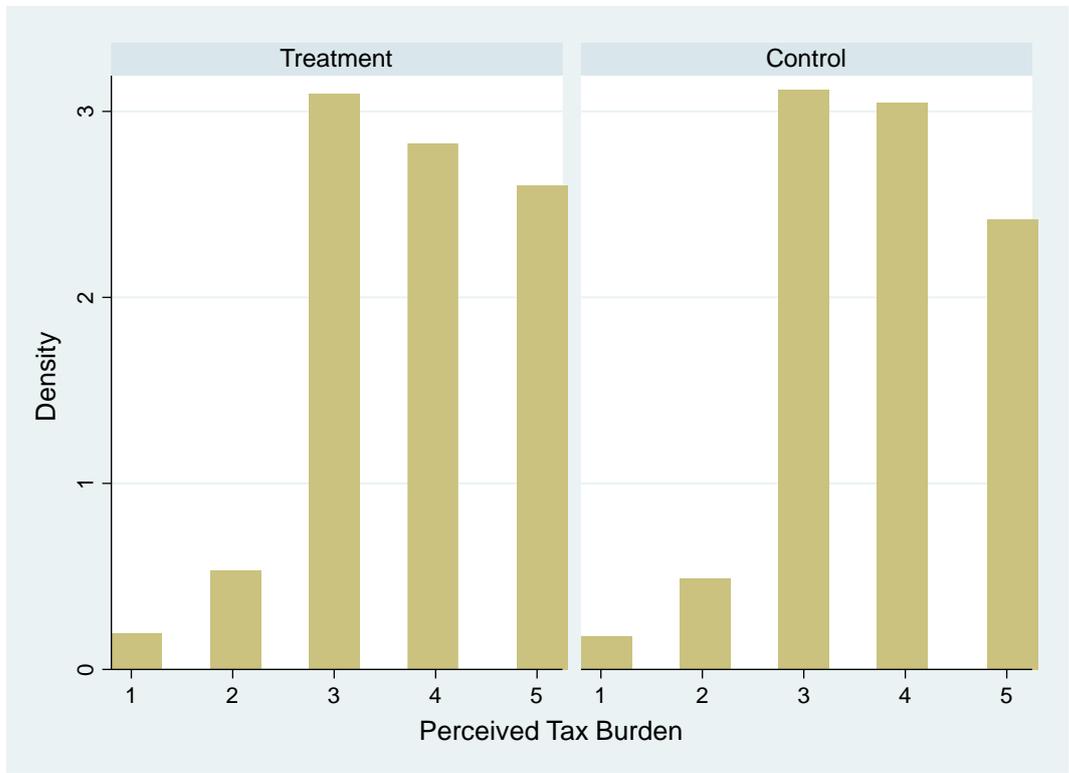

Fig. 1. Histogram of perceived tax burden in the first survey (before correct information about income position).

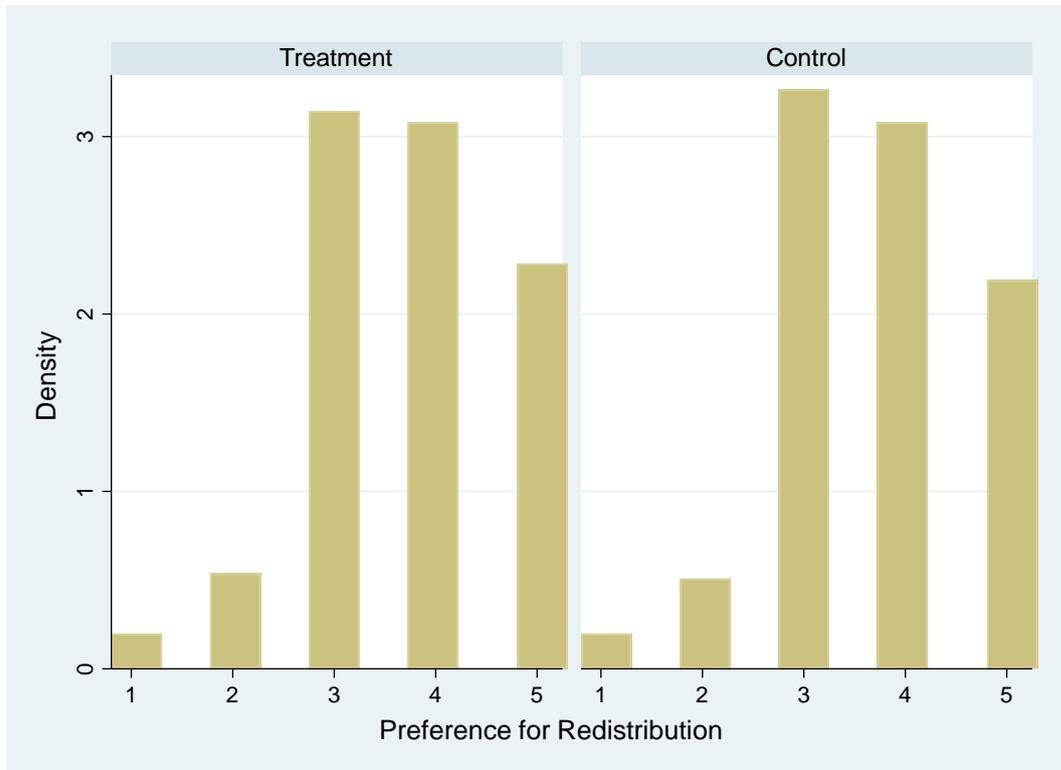

Fig. 2. Histogram of preference for redistribution in the first survey (before correct information about income position).

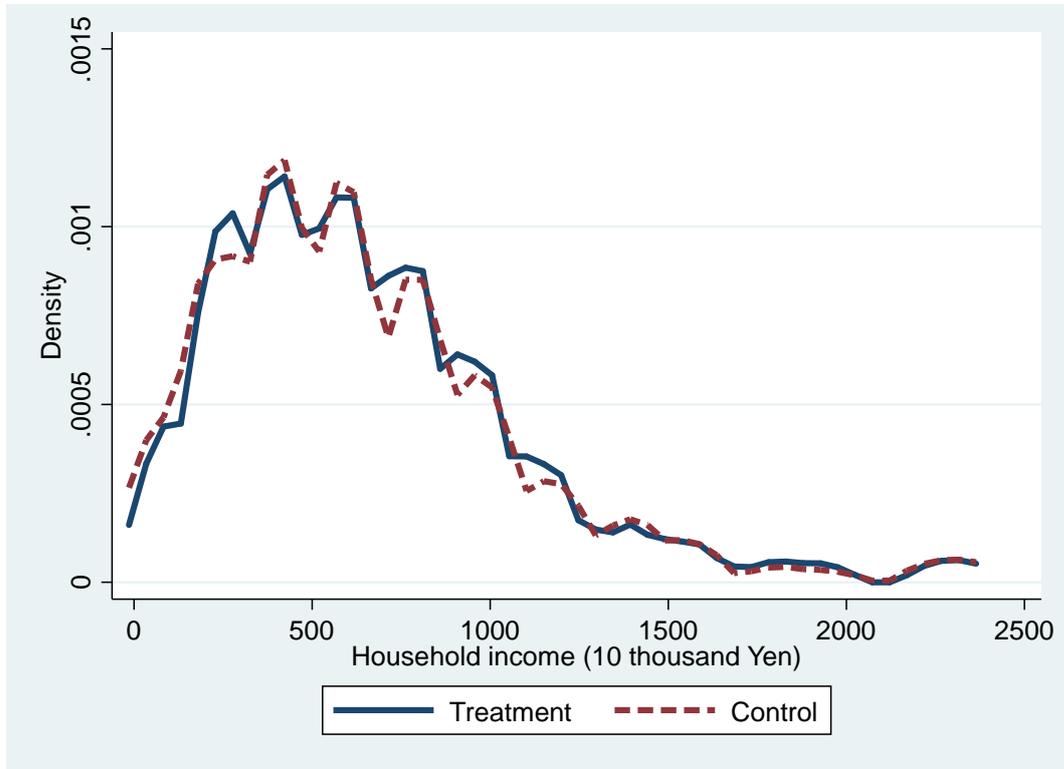

Fig. 3. Kernel density for household income.

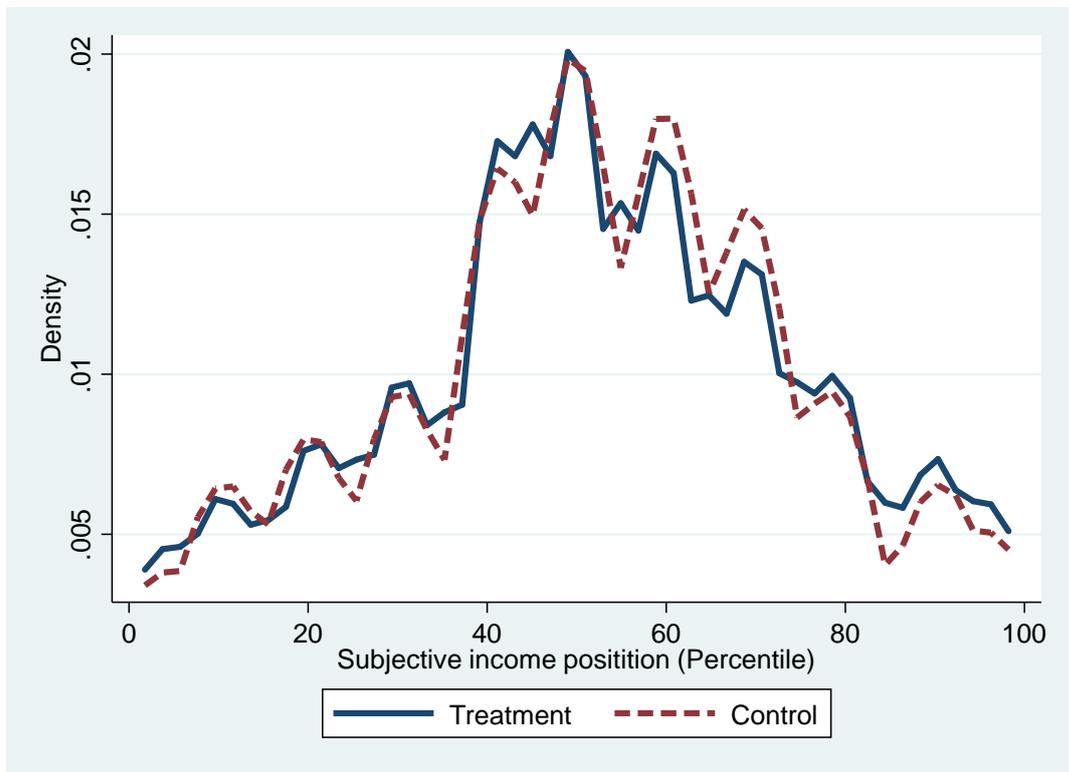

Fig. 4. Kernel density of subjective income position.

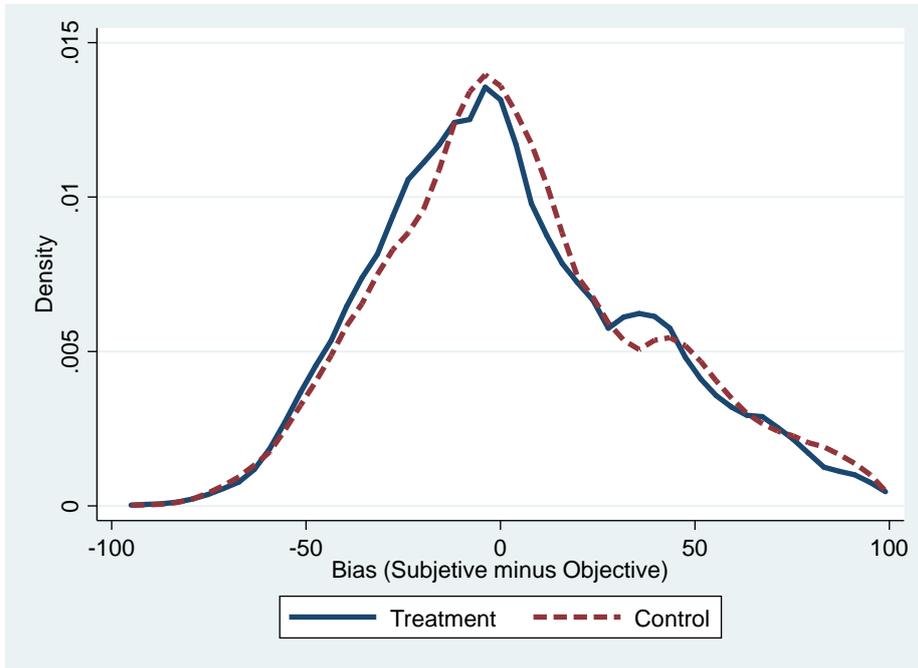

Fig. 5 (a). Kernel density for biased subjective income position (difference between subjective income position and income position in the residential prefecture).

Note: The value is "income position in the same residential prefecture" minus "subjective income position" in the first survey. Negative value indicates that the individual's subjective income position is lower than the real income position in the residential prefecture.

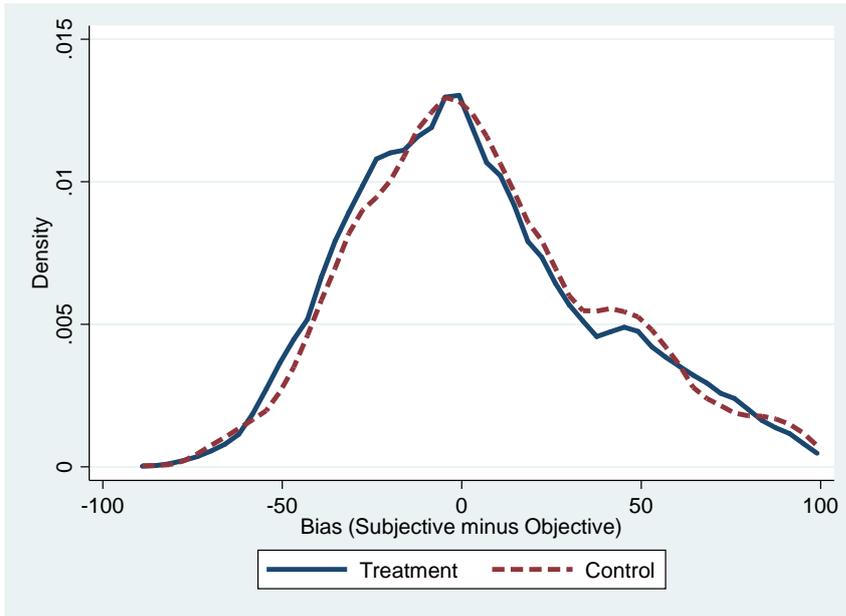

Fig. 5 (b). Kernel density for biased subjective income position in the first survey (difference between subjective income position and income position in the age group)
Note: The value is "income position in the same cohort" minus "subjective income position" in the first survey. Negative value indicates that the individual's subjective income position is lower than the real income position within the cohort.

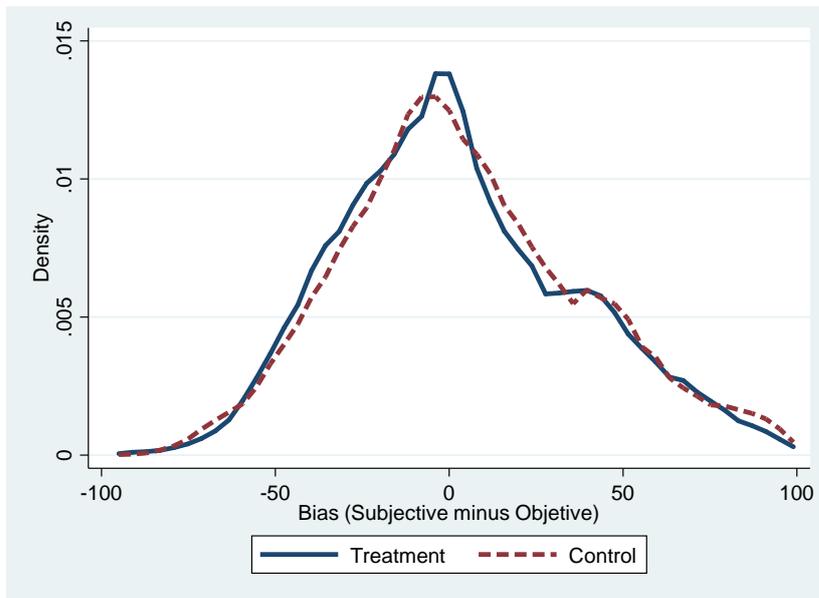

Fig. 5 (c). Kernel density for biased subjective income position in the first survey (difference between subjective income position and income position in the educational background).
Note: The value is "income position in the same educational background" minus "subjective income position." Negative value indicates that the individual's subjective income position is lower than the real income position in the educational background.

Table 1. Balance check.

Mean difference test between treatment and control groups in the first survey.

| | Description | (1) Treatment | (2) Control | Absolute t-values |
|---|---|---|---|---|
| TAX BURDEN | Degree of perceived tax burden : 1 (strongly disagree) – 5 (strongly agree). | 3.78 | 3.78 | 0.34 |
| PPREF_REDIST | Degree of agreement with the argument that the government should reduce income inequality: 1 (strongly disagree) – 5 (strongly agree). | 3.73 | 3.72 | 0.30 |
| SUB_INC | Subjective income position in society in the first wave (1–100%). Larger value is higher position. | 51.8 | 52.0 | 0.28 |
| BIAS_LOCAL | SUB_INC – Income position in the same residential prefecture. Larger value indicates that subjective income position is higher than the real one. | −2.84 | −4.72 | 1.51 |
| BIAS_AGE | SUB_INC – Income position in the same cohort. Larger value indicates that subjective income position is higher than the real one. | −3.76 | −5.71 | 1.67* |
| BIAS_EDU | SUB_INC – Income position in the same educational background. Larger value indicates that subjective income position is higher than the real one. | −2.56 | −4.55 | 1.45 |
| POS_BIAS_LOCAL | BIAS_LOCAL if BIAS_LOCAL>0, otherwise 0. | 13.1 | 12.3 | 1.29 |
| POS_BIAS_AGE | BIAS_AGE if BIAS_AGE>0, otherwise 0. | 12.5 | 11.3 | 1.95* |
| POS_BIAS_EDU | BIAS_EDU if BIAS_EDU>0, otherwise 0. | 16.3 | 17.3 | 1.14 |
| NEG_BIAS_LOCAL | BIAS_LOCAL if BIAS_LOCAL<0, otherwise 0. | −15.9 | −16.9 | 1.18 |
| NEG_BIAS_AGE | BIAS_AGE if BIAS_AGE<0, otherwise 0. | −15.9 | −16.9 | 1.22 |
| NEG_BIAS_EDU | BIAS_EDU if BIAS_EDU<0, otherwise 0. | −12.9 | −11.9 | 1.57 |
| HINCOM | Household income (10 thousand yen). | 656.0 | 639.6 | 1.00 |
| GINI | It is Gini coefficient of respondent's residential area, otherwise 0. | 0.33 | 0.33 | 1.30 |
| AGE | Ages. | 44.9 | 44.5 | 0.93 |
| UUNIV | It has 1 if respondent graduated from university or more, otherwise 0. | 0.24 | 0.24 | 0.14 |
| NOJOB | It has 1 if respondent does not have job, otherwise 0. | 0.07 | 0.09 | 1.20 |
| CHILD | Number of children. | 0.87 | 0.88 | 0.12 |
| MARRI | It has 1 if respondent is married, otherwise 0. | 0.55 | 0.56 | 0.62 |
| IRRECIPRO | If someone does me a favor, I am prepared to return it 1 (strongly agree) – 5 (strongly disagree). | 1.96 | 1.94 | 0.72 |
| IRRECIPRO_D | It has 1 if respondent chose 3, 4, or 5 for IRRECIPRO, otherwise 0. | 0.24 | 0.22 | 1.26 |
| COMPET | Competition results in benefits for society 1 (strongly disagree) – 5 (strongly agree). | 3.69 | 3.66 | 1.16 |
| COMPET_D | It has 1 if respondent chose 3, 4, or 5 for COMPET, otherwise 0. | 0.94 | 0.92 | 2.31** |
| Observations | | 2,341 | 1,134 | |

Notes: Each value is standardized for comparison. * suggests statistical significance at the 10% level

Table 2. Estimation for bias (OLS model)
Full-sample

|  | Bias | | | Absolute value | | |
|---|---|---|---|---|---|---|
|  | (1) BIAS_LOCAL | (2) BIAS_AGE | (3) BIAS_EDU | (4) BIAS_LOCAL | (5) BIAS_AGE | (6) BIAS_EDU |
| IRRECIPRO | 3.192*** | 2.745*** | 3.042*** | 1.231** | 1.062** | 0.862* |
|  | (0.61) | (0.61) | (0.63) | (0.49) | (0.49) | (0.49) |
| COMPET | −1.308** | −2.146*** | −1.434** | −1.779*** | −1.734*** | −1.837*** |
|  | (0.63) | (0.63) | (0.66) | (0.52) | (0.53) | (0.53) |
| UNIV | 0.65 | −1.119 | 6.772*** | −1.528** | −2.076** | −1.979** |
|  | (1.07) | (1.04) | (1.05) | (0.86) | (0.85) | (0.85) |
| HINCOM | −0.049*** | −0.051*** | −0.049*** | −0.004*** | −0.004*** | −0.003*** |
|  | (0.001) | (0.001) | (0.001) | (0.001) | (0.001) | (0.001) |
| GINI | −7.181 | −1.293 | 5.422 | 0.275 | 15.6 | 4.83 |
|  | (19.1) | (17.9) | (18.1) | (14.6) | (14.4) | (14.3) |
| AGE | −0.015 | 0.069* | 0.021 | 0.025 | 0.035 | −0.010 |
|  | (0.04) | (0.41) | (0.04) | (0.03) | (0.03) | (0.03) |
| CHILD | −0.909* | −0.411 | −0.354 | −0.574 | −0.598 | −0.411 |
|  | (0.47) | (0.47) | (0.47) | (0.37) | (0.38) | (0.37) |
| TREAT DUMMY | −1.099 | −1.268 | −0.926 | 0.292 | 0.352 | −0.310 |
|  | (0.95) | (0.93) | (0.96) | (0.77) | (0.77) | (0.77) |
| Constant | 57.75*** | 55.56*** | 49.75*** | 40.54*** | 34.96*** | 33.15*** |
|  | (7.00) | (6.67) | (6.72) | (5.41) | (5.37) | (5.61) |
| Adj- $R^2$ | 0.40 | 0.44 | 0.40 | 0.02 | 0.02 | 0.02 |
| Observations | 3,475 | 3,475 | 3,475 | 3,475 | 3,475 | 3,475 |

Note: *, **, and *** suggest statistical significance at the 10%, 5%, and 1% levels, respectively. The numbers in parentheses are robust standard errors.

Table 3. Effects of biased subjective income position on perceived tax burden (Fixed effects model)

| | (1) Treatment | (2) Control | (3) Treatment | (4) Control | (5) Treatment | (6) Control |
|---|---|---|---|---|---|---|
| SECOND* BIAS_LOCAL | 0.190*** (0.07) | 0.036 (0.10) | | | | |
| SECOND* BIAS_AGE | | | 0.262*** (0.07) | 0.046 (0.11) | | |
| SECOND* BIAS_EDU | | | | | 0.259*** (0.08) | 0.082 (0.108) |
| SECOND* HINCOM | 0.007 (0.01) | 0.006 (0.01) | 0.012* (0.06) | 0.005 (0.01) | 0.010* (0.06) | −0.003 (0.008) |
| SECOND* GINI | −0.349 (0.76) | −0.031 (1.16) | −0.363 (0.76) | −0.031 (1.16) | −0.416 (0.76) | −0.023 (1.16) |
| SECOND* AGE | −0.215 (0.17) | −0.149 (0.27) | −0.242 (0.17) | −0.151 (0.27) | −0.267 (0.17) | −0.149 (0.27) |
| SECOND* CHILD | −0.463 (2.03) | −0.421 (3.04) | −0.461 (2.03) | −0.457 (3.00) | −0.516 (2.03) | −0.511 (3.00) |
| SECOND* UNIVE | 0.459 (4.81) | −6.58 (6.63) | 0.012 (4.82) | −6.48 (6.66) | −2.85 (4.83) | −7.19 (6.68) |
| SECOND | 0.178 (0.27) | 0.244 (0.41) | 0.162 (0.27) | 0.239 (0.41) | 0.211 (0.27) | 0.227 (0.40) |
| Adj- $R^2$ | 0.56 | 0.54 | 0.56 | 0.54 | 0.56 | 0.54 |
| Group | 2,341 | 1,134 | 2,341 | 1,134 | 2,341 | 1,134 |
| Observations | 4,682 | 2,268 | 4,682 | 2,268 | 4,682 | 2,268 |

Note: For convenience of readers, value are multiplied by 100.
*, **, and *** suggest statistical significance at the 10%, 5%, and 1% levels, respectively. The numbers in parentheses are robust standard errors. Time invariant individual characteristics are controlled by the Fixed effects model.
For convenience of readers, with the exception of SECOND*GINI and SECOND, values are multiplied by 100.

Table 4. Effects of biased subjective income position on perceived tax burden (Fixed effects model)

|  | (1) Treatment | (2) Control | (3) Treatment | (4) Control | (5) Treatment | (6) Control |
|---|---|---|---|---|---|---|
| SECOND* POS_BIAS_LOCAL | 0.392*** (0.11) | 0.108 (0.16) |  |  |  |  |
| SECOND* NEG_BIAS_LOCAL | −0.186 (0.16) | −0.065 (0.20) |  |  |  |  |
| SECOND* POS_BIAS_AGE |  |  | 0.308*** (0.10) | 0.052 (0.16) |  |  |
| SECOND* NEG_BIAS_AGE |  |  | 0.159 (0.17) | 0.142 (0.23) |  |  |
| SECOND* POS_BIAS_EDU |  |  |  |  | 0.362*** (0.11) | 0.127 (0.16) |
| SECOND* NEG_BIAS_EDU |  |  |  |  | 0.089 (0.16) | 0.049 (0.20) |
| Adj- $R^2$ | 0.55 | 0.53 | 0.56 | 0.53 | 0.55 | 0.54 |
| Group | 2,209 | 1,060 | 2,226 | 1,075 | 2,154 | 1,054 |
| Observations | 4,418 | 2,120 | 4,452 | 2,150 | 4,308 | 2,108 |

Note:
*, **, and *** suggest statistical significance at the 10%, 5%, and 1% levels, respectively. The numbers in parentheses are robust standard errors. Time invariant individual characteristics are controlled by the Fixed effects model.
For convenience of readers, values are multiplied by 100.

Table 5. Effects of interaction between the information provision and subjective view on perceived tax burden (Fixed effects model)

|  | (1) Treatment | (2) Control | (3) Treatment | (4) Control | (5) Treatment | (6) Control | (7) Treatment | (8) Control |
|---|---|---|---|---|---|---|---|---|
| SECOND* IRRECIPRO | 0.124*** (0.04) | 0.052 (0.06) |  |  |  |  |  |  |
| SECOND* IRRECIPRO_D |  |  | 0.211*** (0.07) | 0.162 (0.12) |  |  |  |  |
| SECOND* COMPET |  |  |  |  | −0.041 (0.04) | −0.048 (0.05) |  |  |
| SECOND* COMPET_D |  |  |  |  |  |  | 0.028 (0.15) | 0.126 (0.17) |
| SECOND* POS_BIAS_LOCAL | 0.336** (0.15) | 0.080 (0.22) | 0.347** (0.152) | 0.078 (0.22) | 0.370** (0.15) | 0.090 (0.22) | 0.397** (0.15) | 0.119 (0.22) |
| SECOND* NEG_BIAS_LOCAL | −0.203 (0.22) | −0.078 (0.28) | −0.201 (0.22) | −0.103 (0.28) | −0.178 (0.22) | −0.061 (0.29) | −0.188 (0.22) | −0.069 (0.29) |
| Adj- $R^2$ | 0.56 | 0.54 | 0.56 | 0.54 | 0.56 | 0.54 | 0.56 | 0.54 |
| Group | 2,209 | 1060 | 2,209 | 1060 | 2,209 | 1060 | 2,209 | 1060 |
| Observations | 4,418 | 2,120 | 4,418 | 2,120 | 4,418 | 2,120 | 4,418 | 2,120 |

Note:
*, **, and *** suggest statistical significance at the 10%, 5%, and 1% levels, respectively. The numbers in parentheses are robust standard errors clustered on presidential prefectures. Time invariant individual characteristics are controlled by the Fixed effects model.
For convenience of readers, values are multiplied by 100 for SECOND*POS_BIAS_LOCAL and NEG_BIAS_LOCAL.

Table 6. Effects of interaction between the information provision and subjective view on perceived tax burden (Fixed effects model)
Sub-sample according to positive and negative biases

Panel A. Treatment group

|  | Positive | | | | Negative | | | |
| --- | --- | --- | --- | --- | --- | --- | --- | --- |
|  | (1) | (2) | (3) | (4) | (5) | (6) | (7) | (8) |
| SECOND* IRRECIPRO | 0.105** (0.05) | | | | 0.134*** (0.04) | | | |
| SECOND* IRRECIPRO_D | | 0.185* (0.09) | | | | 0.242** (0.10) | | |
| SECOND* COMPET | | | 0.008 (0.05) | | | | −0.096* (0.05) | |
| SECOND* COMPET_D | | | | 0.151 (0.21) | | | | −0.197 (0.20) |
| Adj- $R^2$ | 0.54 | 0.54 | 0.54 | 0.53 | 0.58 | 0.58 | 0.58 | 0.57 |
| Group | 1,047 | 1,047 | 1,047 | 1,047 | 1,162 | 1,162 | 1,162 | 1,162 |
| Observations | 2,094 | 2,094 | 2,094 | 2,094 | 2,324 | 2,324 | 2,324 | 2,324 |

Panel B. Control group

|  | Positive | | | | Negative | | | |
| --- | --- | --- | --- | --- | --- | --- | --- | --- |
|  | (1) | (2) | (3) | (4) | (5) | (6) | (7) | (8) |
| SECOND* IRRECIPRO | 0.053 (0.08) | | | | 0.049 (0.08) | | | |
| SECOND* IRRECIPRO_D | | 0.151 (0.15) | | | | 0.190 (0.18) | | |
| SECOND* COMPET | | | −0.037 (0.08) | | | | −0.058 (0.67) | |
| SECOND* COMPET_D | | | | 0.024 (0.24) | | | | 0.323 (0.20) |
| Adj- $R^2$ | 0.56 | 0.54 | 0.56 | 0.52 | 0.56 | 0.56 | 0.56 | 0.56 |
| Group | 536 | 536 | 536 | 536 | 524 | 524 | 524 | 524 |
| Observations | 1,072 | 1,072 | 1,072 | 1,072 | 1,048 | 1,048 | 1,048 | 1,048 |

Note: *, **, and *** suggest statistical significance at the 10%, 5%, and 1% levels, respectively. The numbers in parentheses are robust standard errors clustered on presidential prefectures. Time invariant individual characteristics are controlled by the Fixed effects model. Independent variables included in

Table 3 are included but not reported.

Table 7. Effects of biased subjective income position on preference for redistribution (Fixed effects model)

|  | (1) Treatment | (2) Control | (3) Treatment | (4) Control | (5) Treatment | (6) Control |
|---|---|---|---|---|---|---|
| SECOND* BIAS_LOCAL | 0.036 (0.07) | 0.152 (0.10) |  |  |  |  |
| SECOND* BIAS_AGE |  |  | 0.036 (0.07) | 0.130 (0.10) |  |  |
| SECOND* BIAS_EDU |  |  |  |  | 0.050 (0.07) | 0.146 (0.09) |
| SECOND* HINCOM | 0.002 (0.006) | 0.005 (0.08) | 0.002 (0.006) | 0.005 (0.08) | 0.001 (0.006) | 0.005 (0.08) |
| SECOND* GINI | −1.39* (0.78) | 1.24 (1.13) | −1.38* (0.78) | 0.003 (1.13) | −1.35* (0.79) | −0.004 (1.13) |
| SECOND* AGE | −0.416** (0.17) | −0.219 (0.23) | −0.419** (0.17) | −0.223 (0.23) | −0.409** (0.17) | −0.218 (0.23) |
| SECOND* CHILD | 2.800 (1.99) | 2.334 (2.83) | 2.792 (1.99) | 2.167 (2.81) | 2.715 (1.99) | 2.127 (2.83) |
| SECOND* UNIVE | −0.195 (4.75) | −1.02 (6.59) | −0.131 (4.72) | −0.715 (6.59) | −0.473 (4.78) | −1.861 (6.65) |
| SECOND | 0.497* (0.27) | 3.90 (38.6) | 0.498* (0.27) | −0.28 (0.39) | 0.479* (0.27) | −0.0.27 (0.39) |
| Adj- $R^2$ | 0.56 | 0.62 | 0.56 | 0.62 | 0.56 | 0.62 |
| Group | 2,341 | 1,134 | 2,341 | 1,134 | 2,341 | 1,134 |
| Observations | 4,682 | 2,268 | 4,682 | 2,268 | 4,682 | 2,268 |

Note:
*, **, and *** suggest statistical significance at the 10%, 5%, and 1% levels, respectively. The numbers in parentheses are robust standard errors clustered on individual players. Time invariant individual characteristics are controlled by the Fixed effects model.
For convenience of readers, with the exception of SECOND*GINI and SECOND, values are multiplied by 100.

Table 8. Effects of biased subjective income position on preference for redistribution (Fixed effects model)

|  | (1) Treatment | (2) Control | (3) Treatment | (4) Control | (5) Treatment | (6) Control |
|---|---|---|---|---|---|---|
| SECOND* POS_BIAS_LOCAL | 0.078 (0.10) | 0.113 (0.14) |  |  |  |  |
| SECOND* NEG_BIAS_LOCAL | −0.102 (0.15) | 0.276 (0.20) |  |  |  |  |
| SECOND* POS_BIAS_AGE |  |  | 0.103 (0.10) | 0.182 (0.13) |  |  |
| SECOND* NEG_BIAS_AGE |  |  | 0.132 (0.16) | 0.184 (0.21) |  |  |
| SECOND* POS_BIAS_EDU |  |  |  |  | 0.198* (0.11) | 0.149 (0.14) |
| SECOND* NEG_BIAS_EDU |  |  |  |  | −0.200 (0.15) | 0.191 (0.19) |
| Adj- $R^2$ | 0.55 | 0.62 | 0.55 | 0.61 | 0.56 | 0.61 |
| Group | 2,341 | 1,134 | 2,341 | 1,134 | 2,341 | 1,134 |
| Observations | 4,682 | 2,268 | 4,682 | 2,268 | 4,682 | 2,268 |

Note:
*, **, and *** suggest statistical significance at the 10%, 5%, and 1% levels, respectively. The numbers in parentheses are robust standard errors. Time invariant individual characteristics are controlled by the Fixed effects model.
For convenience of readers, values are multiplied by 100.

Table 9. Effects of interaction between the information provision and subjective view on preference for redistribution (Fixed effects model)

|  | (1) Treatment | (2) Control | (3) Treatment | (4) Control | (5) Treatment | (6) Control | (7) Treatment | (8) Control |
|---|---|---|---|---|---|---|---|---|
| SECOND* IRRECIPRO | 0.128*** (0.04) | 0.063 (0.05) | | | | | | |
| SECOND* IRRECIPRO_D | | | 0.201*** (0.06) | 0.138 (0.10) | | | | |
| SECOND* COMPET | | | | | −0.091** (0.04) | −0.086* (0.05) | | |
| SECOND* COMPET_D | | | | | | | −0.124 (0.13) | −0.050 (0.17) |
| SECOND* POS_BIAS_LOCAL | 0.020 (0.14) | 0.079 (0.19) | 0.035 (0.15) | 0.087 (0.19) | 0.029 (0.15) | 0.081 (0.19) | 0.058 (0.14) | 0.108 (0.19) |
| SECOND* NEG_BIAS_LOCAL | −0.119 (0.21) | 0.260 (0.27) | −0.118 (0.20) | 0.243 (0.28) | −0.087 (0.20) | −0.284 (0.28) | −0.092 (0.21) | 0.278 (0.28) |
| Adj- $R^2$ | 0.56 | 0.61 | 0.56 | 0.61 | 0.56 | 0.61 | 0.56 | 0.54 |
| Group | 2,209 | 1,060 | 2,209 | 1,060 | 2,209 | 1,060 | 2,209 | 1060 |
| Observations | 4,418 | 2,120 | 4,418 | 2,120 | 4,418 | 2,120 | 4,418 | 2,120 |

Note:
*, **, and *** suggest statistical significance at the 10%, 5%, and 1% levels, respectively. The numbers in parentheses are robust standard errors clustered on presidential prefectures. Time invariant individual characteristics are controlled by the fixed effects model.
For convenience of readers, values are multiplied by 100 for SECOND*POS_BIAS_LOCAL and NEG_BIAS_LOCAL.

Table 10. Effects of interaction between the information provision and subjective view on preference for redistribution (Fixed effects model)

Sub-sample according to positive and negative biases

Panel A. Treatment group

|  | Positive | | | | Negative | | | |
| --- | --- | --- | --- | --- | --- | --- | --- | --- |
|  | (1) | (2) | (3) | (4) | (5) | (6) | (7) | (8) |
| SECOND* IRRECIPRO | 0.146*** (0.04) | | | | 0.111** (0.05) | | | |
| SECOND* IRRECIPRO_D | | 0.240*** (0.09) | | | | 0.158 (0.10) | | |
| SECOND* COMPET | | | −0.092* (0.05) | | | | −0.093* (0.05) | |
| SECOND* COMPET_D | | | | −0.119 (0.19) | | | | −0.146 (0.18) |
| Adj-$R^2$ | 0.56 | 0.56 | 0.56 | 0.56 | 0.56 | 0.56 | 0.56 | 0.56 |
| Group | 1,047 | 1,047 | 1,047 | 1,047 | 1,162 | 1,162 | 1,162 | 1,162 |
| Observations | 2,094 | 2,094 | 2,094 | 2,094 | 2,324 | 2,324 | 2,324 | 2,324 |

Panel B. Control group

|  | Positive | | | | Negative | | | |
| --- | --- | --- | --- | --- | --- | --- | --- | --- |
|  | (1) | (2) | (3) | (4) | (5) | (6) | (7) | (8) |
| SECOND* IRRECIPRO | 0.051 (0.07) | | | | 0.079 (0.07) | | | |
| SECOND* IRRECIPRO_D | | 0.084 (0.13) | | | | −0.213 (0.146) | | |
| SECOND* COMPET | | | −0.055 (0.07) | | | | −0.118* (0.07) | |
| SECOND* COMPET_D | | | | −0.012 (0.22) | | | | −0.098 (0.27) |
| Adj-$R^2$ | 0.54 | 0.54 | 0.54 | 0.54 | 0.67 | 0.68 | 0.67 | 0.67 |
| Group | 536 | 536 | 536 | 536 | 524 | 524 | 524 | 524 |
| Observations | 1,072 | 1,072 | 1,072 | 1,072 | 1,048 | 1,048 | 1,048 | 1,048 |

Note: *, **, and *** suggest statistical significance at the 10%, 5%, and 1% levels, respectively. The numbers in parentheses are robust standard errors

clustered on presidential prefectures. Time invariant individual characteristics are controlled by the fixed effects model. Independent variables included in Table 7 are included but not reported.

# Appendix:

## Sampling method

During October 25–30, 2018, I commissioned the Nikkei Research Company to conduct a nationally representative web survey covering all parts of Japan. Prior to surveys, the company had the basic information of subjects (residential areas, ages, etc.). This was because subjects were obliged to answer basic questions about themselves when they were registered in the list of subjects. Hence, the questionnaire can be sent to 9,130 subjects who are designed to be representative of the Japanese population in terms of residential areas, ages, and genders.

In the questionnaire, there are various characteristics of respondents, such as economic condition, demographic information, educational background, subjective views, and perception about government policy. Internet users are presumably different from non-internet users. However, according to an official survey on information technology, nearly 100% of Japanese people in the 20–29, 30–39, and 40–49 years age groups were internet users in 2015. Even for older cohorts, the percentage of web users is over 90% for people aged 50–59 years and 80% for people aged 60–69 years. Hence, the sampling method through the Internet is unlikely to suffer bias.

After collecting information about household income, individual's income positions are calculated in three ways:

(1) In Japan, an administrative subdivision is called a prefecture, and there are 47 prefectures. Residential prefectures are used as residential locality. Subsequently, within a residential prefecture, an individual's income position is calculated.

(2) Respondents' ages are in the range 19–68 years. I divided respondents into every ten years cohort (19–29, 30–39, 40-49, 50–59, and 60–69 years age groups). For instance, an individual's income position is calculated in the 30–40 years age group if they belong to that group.

(3) Respondents' final educational backgrounds comprise i) junior high school, ii) high school, iii) vocational training school, iv) college, and v) university or graduate school. For instance, an individual's income position is calculated in the group with high school as final educational background if their final educational background is high school.

Table A1. Confidence in government: (%)

| | |
|---|---|
| Sweden | 59.9 |
| South Korea | 49.5 |
| New Zealand | 44.7 |
| Germany | 44.4 |
| Netherland | 33.0 |
| United States | 32.6 |
| Argentina | 31.6 |
| Australia | 30.0 |
| Japan | 24.3 |

Source: World Value Survey 2010–2014.

Percentage of positive responses ("A great deal" or "Quite a lot")

Question: Could you tell me how much confidence you have in the Government: is it a great deal of confidence, quite a lot of confidence, not very much confidence, or none at all?

Selected samples: Argentina 2013, Australia 2012, Germany 2013, Japan 2010, Netherlands 2012, New Zealand 2011, South Korea 2010, Spain 2011, Sweden 2011, United States 2011

Table A 2. Estimation for perceived income rank (SUB_INC_) (OLS model) Full-sample

|  | (1) | (2) | (3) |
|---|---|---|---|
| Objective income position | 0.014 (0.05) | −0.018 (0.06) | 0.072 (0.05) |
| Objective Income position _LOCAL | −0.028 (0.03) | | |
| Objective Income position _AGE | | 0.008 (0.04) | |
| Objective Income position _EDU | | | −0.100*** (0.03) |
| Adj- $R^2$ | 0.04 | 0.04 | 0.04 |
| Observations | 3,475 | 3,475 | 3,475 |

Note: *, **, and *** suggest statistical significance at the 10%, 5%, and 1% levels, respectively. Control variables used in Table 2 are included in the model but are not reported. The numbers in parentheses are robust standard errors.

Fig. A1. Question about perceived income position in the first wave.

After answering questions including the perceived tax burden and preference for redistribution as above, respondents were asked about their subjective income position in society.

Fig. A2. For the treatment group, true relative income position calculated based on the first survey is informed before starting the survey.

次の情報を参考にして、質問にお答えください。

10月に実施しました「子供時代の生活環境と生活意識に関する調査」で、あなたにご回答いただいた所得などのデータを用い、全体結果から算出した**あなたの所得地位**は下記のようになります。

同一学歴のグループの中でのあなたの所得地位（全体の中の上位$Aval1%）

同一年齢のグループの中でのあなたの所得地位（全体の中の上位$Aval2%）

同一県内のグループの中でのあなたの所得地位（全体の中の上位$Aval3%）

Directly after the information, respondents asked about the questions including the perceived tax burden and preference for redistribution as below, although the number of questions reduced to put focus.

**Q18A**

日本の政治や政策、企業の責任について、あなたのご意見をお伺いします。 （それぞれひとつずつ）

| | 同意する | やや同意する | どちらともいえない | あまり同意しない | 同意しない |
|---|---|---|---|---|---|
| 政府は貧富の差を小さくすべきである | (q18a_14)=5 | (q18a_14)=4 | (q18a_14)=3 | (q18a_14)=2 | (q18a_14)=1 |
| 所得格差は深刻な問題である | (q18a_27)=5 | (q18a_27)=4 | (q18a_27)=3 | (q18a_27)=2 | (q18a_27)=1 |
| 最低賃金を引き上げるべきである | (q18a_28)=5 | (q18a_28)=4 | (q18a_28)=3 | (q18a_28)=2 | (q18a_28)=1 |
| 自分自身の税負担を重いと感じる | (q18a_30)=5 | (q18a_30)=4 | (q18a_30)=3 | (q18a_30)=2 | (q18a_30)=1 |

Fig. A3. Distribution of Irreciprocity (IRRECIP)

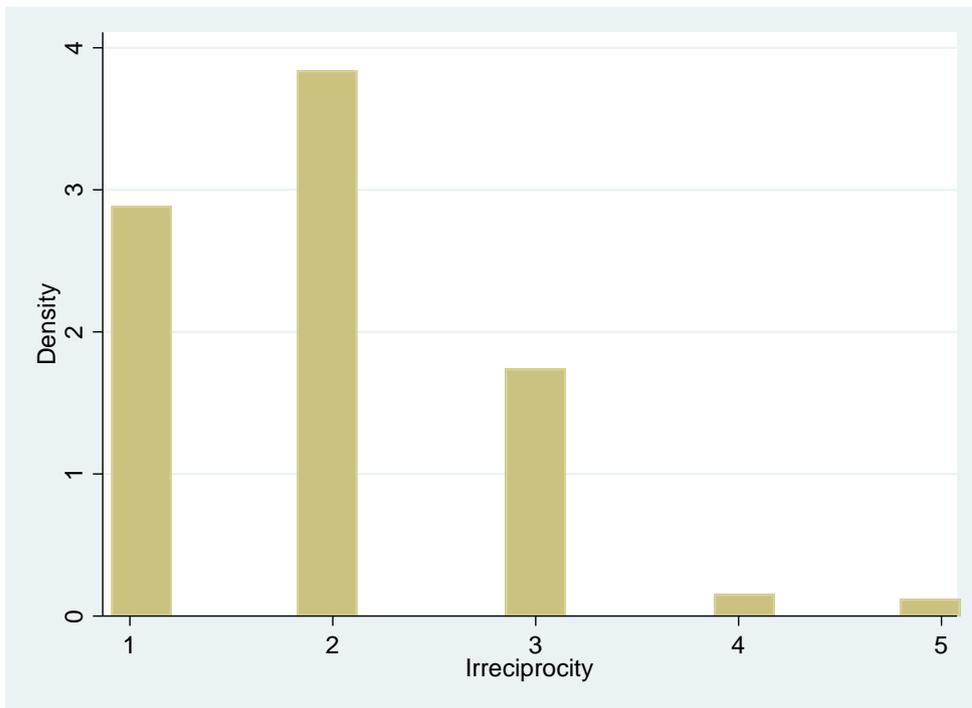

Fig. A4. Distribution of View about competition (COMPET)

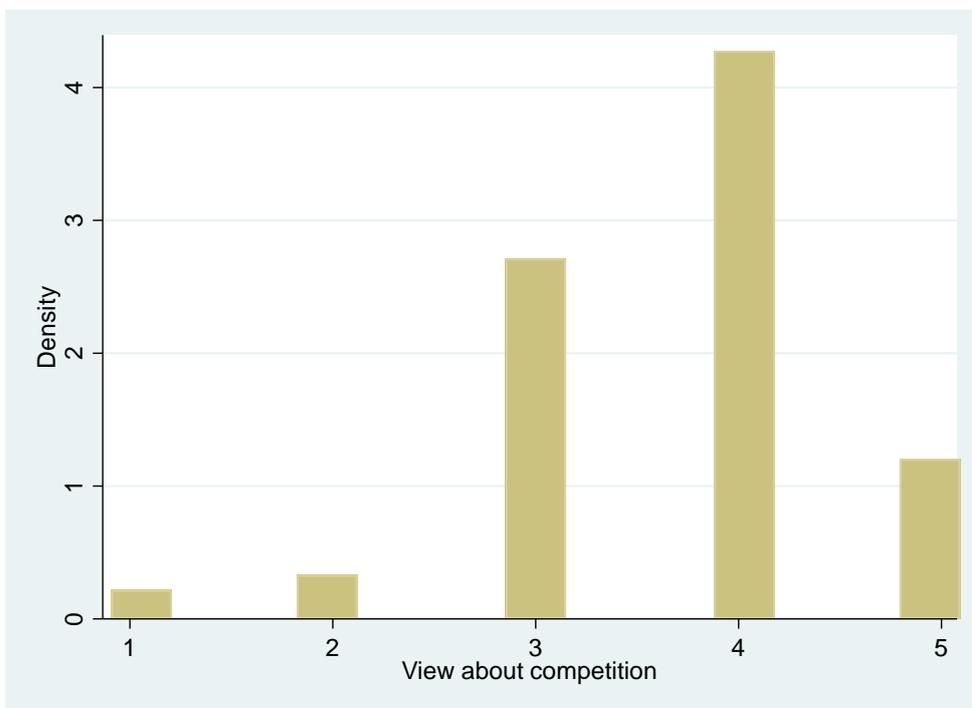